\pdfoutput=1
\documentclass[twocolumn,
superscriptaddress,
prl,
showpacs,
preprintnumbers,
amsmath,amssymb]{revtex4}

\usepackage{graphicx,epsf,epsfig,ulem}
\usepackage{epsfig}
\usepackage{epstopdf} 

\usepackage{bm}
\usepackage{subfigure}
\usepackage{color}

\begin{document}

\title{Spin-lattice relaxation times of single donors and donor clusters in silicon}

\author{Yuling Hsueh*}
\affiliation{Network for Computational Nanotechnology, Purdue University, West Lafayette, IN 47907, USA}

\author{Holger B\"{u}ch}
\affiliation{Centre for Quantum Computation and Communication Technology, School of Physics, University of New South Wales, Sydney, NSW 2052, Australia}

\author{Yaohua Tan}
\affiliation{Network for Computational Nanotechnology, Purdue University, West Lafayette, IN 47907, USA}

\author{Yu Wang}
\affiliation{Network for Computational Nanotechnology, Purdue University, West Lafayette, IN 47907, USA}

\author{Lloyd C. L. Hollenberg}
\affiliation{Centre for Quantum Computation and Communication Technology, School of Physics, University of Melbourne, VIC 3010, Australia}

\author{Gerhard Klimeck}
\affiliation{Network for Computational Nanotechnology, Purdue University, West Lafayette, IN 47907, USA}

\author{Michelle Y. Simmons}
\affiliation{Centre for Quantum Computation and Communication Technology, School of Physics, University of New South Wales, Sydney, NSW 2052, Australia}

\author{Rajib Rahman*}
\affiliation{Network for Computational Nanotechnology, Purdue University, West Lafayette, IN 47907, USA}

\date{\today}

\begin{abstract}

An atomistic method of calculating the spin-lattice relaxation times ($T_1$) is presented for donors in silicon nanostructures comprising of millions of atoms. The method takes into account the full band structure of silicon including the spin-orbit interaction. 
The electron-phonon Hamiltonian, and hence the deformation potential, is directly evaluated from the strain-dependent tight-binding Hamiltonian. The technique is applied to single donors and donor clusters in silicon, and explains the variation of $T_1$ with the number of donors and electrons, as well as donor locations. Without any adjustable parameters, the relaxation rates in a magnetic field for both systems are found to vary as $B^5$ in excellent quantitative agreement with experimental measurements. The results also show that by engineering electronic wavefunctions in nanostructures, $T_1$ times can be varied by orders of magnitude.  

\end{abstract}


\maketitle 

Due to the extremely long spin coherence times, in some cases exceeding seconds \cite{Lyon_nature_mat,Thewalt_science}, and the existing industrial fabrication infrastructure, silicon is well-suited to be an outstanding platform for semiconductor quantum computer technology \cite{Kane,Hollenberg_2006,deSousa,Calderon,Friesen,Charles}. Qubits hosted by donors in silicon \cite{Kane} have some added advantages as they are readily available few-electron systems with a rich electronic structure and can form identical qubits \cite{RMP_Lloyd}. In the last few years, several key experimental milestones have been achieved in dopant based quantum computing, including the demonstration of electron \cite{Morello_equbit} and nuclear \cite{Morello_nqubit} spin qubits, single spin read-out and initialization \cite{Morello_single_shot,Holger_Ncomm}, and the observation of spin blockade and exchange towards two qubit coupling \cite{Bent_dqd_spin_blockade,Bent_new}. Recent advances in Scanning Tunneling Microscope (STM) lithography has enabled placement of single donors with atomic scale precision \cite{Schofield_PRL}, with the result that various functional units such as quantum wires \cite{Bent_wire}, single electron transistors (SET) \cite{Holger_Ncomm,Martin_single_atom}, and quantum dots \cite{Martin_dot} can all be realized in-plane with densely packed donor islands. The STM approach provides the fabrication precision needed to develop test-bed quantum chips for the demonstration of quantum algorithms in a solid-state quantum computer. 

One of the two most important timescales for a spin qubit is the spin-lattice relaxation time ($T_1$). Recent experiments have measured $T_1$ times in a single donor and in a few-donor cluster indicating shorter $T_1$ times in the latter \cite{Morello_single_shot,Holger_Ncomm}. Previous theoretical works exist in the literature qualitatively describing two different spin relaxation mechanisms in a bulk donor system \cite{Hasegawa,Roth}. However, a comprehensive quantitative theory which combines all the different mechanisms under a unified framework and accounts for the local inhomogeneous environment of the donors in a realistic nanostructure is still lacking. Moreover, there is no theoretical work yet to explain the measured $T_1$ times in densely packed donor clusters. In this letter, we present a comprehensive approach to compute the $T_1$ times in single donors and donor clusters in silicon nanostructures based on the self-consistent atomistic tight-binding (TB) method. The computed $T_1$ times can explain the experimental results of Refs \cite{Morello_single_shot,Holger_Ncomm} without any adjustable parameters. The $T_1$ times are found to depend strongly on the size and shape of the electronic wavefunctions, which suggests that quantum confinement plays an important role in the relaxation process. The calculations also provide an insight into how the $T_1$ times can be engineered by several orders of magnitude in these quantum devices.  

\vspace{-0.25cm}
  
\begin{figure}[htbp]
\includegraphics[trim=50 0 50 0, clip, width=\linewidth]{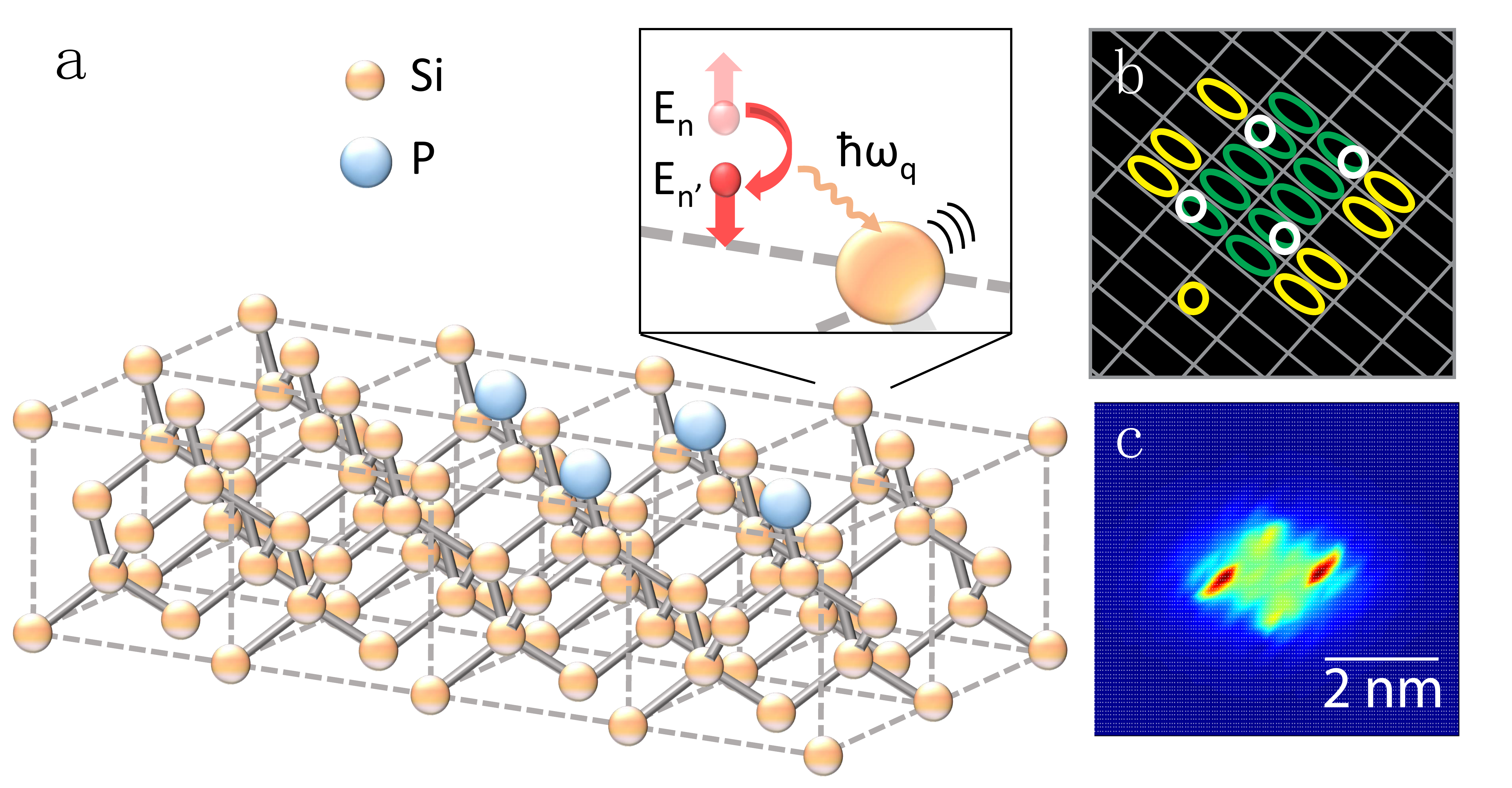}
\caption{a) Schematic plot of a 4P donor cluster in silicon. The cluster hosts an up spin electron with energy $E_n$ which relaxes to the down spin state $E_{n'}$ by emitting a phonon with energy $\hbar\omega_q$. b) An STM template of a 4P cluster showing example dimer locations \cite{Holger_Ncomm}. c) The computed self-consistent probability density of the outermost electron in a 4P donor cluster with 5 electrons (4P5e).}
\vspace{-0.5cm}
\label{fi0}
\end{figure}

Single shot measurements of the donor-bound electron spin performed in ion-implanted metal-oxide-semiconductor (MOS) devices yielded a $T_1$ time of 2.3 s at $B=2$ T and a $B^5$ dependency of the relaxation rate ($1/T_1$) \cite{Morello_single_shot}. Similar $T_1$ measurements have since been performed in a STM patterned donor cluster in an all epitaxially engineered device with atomic scale precision, giving $T_1$ time of 0.4 s at $B=2$ T and a similar $B^5$ dependency of the relaxation rate \cite{Holger_Ncomm}. The donor clusters are composed of several donors with an average separation less than or equal to the single donor Bohr radii (Fig. 1b) which can host several electrons. Such a cluster provides for additional addressability within an array of qubits \cite{Holger_Ncomm}. In general, the $T_1$ times of donors in realistic devices and nanostructures are likely to be influenced by the inhomogeneous environment, and need to be understood from an atomic scale theory.           

The full TB Hamiltonian of the silicon and the P atoms was represented with a 20-orbital $sp^3d^5s^*$ basis per atom including nearest-neighbor and spin-orbit interactions \cite{Klimeck_ted, CMES, Slater-Koster}. A donor was represented by a Coulomb potential of a positive charge screened by the dielectric constant of Si and subjected to an onsite cut off potential \cite{Rahman_prl, Shaikh}. The model reproduces the full energy spectrum of a single donor including valley-orbit splitting \cite{Rahman_orbital_stark_effect}, and also captures the single donor hyperfine \cite{Rahman_prl} and spin-orbit Stark effects \cite{Rahman_spin_orbit} in close agreement with experiments. 
Recent STM imaging of the donor wavefunction also shows excellent agreement with the tight-binding wavefunction at the atomic scale \cite{Salfi}.   
The magnetic field is represented by a vector potential in a symmetric gauge and entered through a Peierls substitution. To capture multi-electron occupation of the donor clusters, a self-consistent Hartree method was employed, in which the electronic charge density was computed from the $N$ lowest energy occupied wavefunctions, $n(r)=\sum_{i}^{N} |\Psi_i(r)|^2$. Solving the potential due to $n(r)$ self-consistently with the tight-binding Hamiltonian until convergence enables us to obtain the binding energies and the wavefunctions of the few electrons bound to the donor cluster. This method has also successfully reproduced the experimental $D^-$ (i.e., the 2nd electron bound to a single P) binding energy \cite{Rahman_ce, Martin_single_atom}. Exchange and correlation terms based on local density approximation are added to the Hartree potential \cite{Lee}. The same methodology has been used to reproduce experimentally measured addition energies of multi-donor clusters \cite{Bent_new}. The TB Hamiltonian of about 1.4 million atoms including the Hartree potential is then solved by a parallel Block Lanczos algorithm to obtain the relevant lowest energy wavefunctions. 

For the relaxation times computed in this work, we assume that an electron has been loaded to the ground state of a single donor or a donor cluster. In case of a bulk P donor, this represents the $A_1$ state at -45.6 meV below the conduction band \cite{Kohn}. This binding energy can vary with donor depths and fields in a realistic device \cite{Rahman_orbital_stark_effect}. In addition, the binding energy in a donor cluster can be sensitive to the donor numbers, electron numbers, and donor locations. Experimentally, the electron can be loaded in the ground state of the donor/cluster by bringing the Fermi level of an electron reservoir in resonance with the Zeeman split ground state, as demonstrated in Refs \cite{Morello_single_shot, Holger_Ncomm}. 

The relaxation rate $1/T_1$, for an electron-phonon interaction Hamiltonian $\hat{H}_{ep}$ can be obtained by Fermi's Golden Rule,

\vspace{-0.5cm}
\begin{equation}
  \frac{1}{T_1}=\frac{2\pi}{\hbar}|\langle{n',n_q+1|\hat{H}_{ep}|n,n_q}\rangle|^2\delta(E_n-E_{n'}-\hbar\omega_q) \label{vgl7}
\end{equation} 
\noindent
where $n$ and $n'$ are the up and down spin electronic states with energy $E_n$ and $E_{n'}$ respectively, $\omega_q$ is the angular frequency of the emitted phonon, $n_q$ and $n_q+1$ are the initial and final phonon states with wavevector $q$. In the B-field range of experimental interest \cite{Morello_single_shot,Holger_Ncomm}, the Zeeman splitting is less than 1 meV. As a result, only acoustic phonons contribute to the spin relaxation process. $\hat{H}_{ep}$ then depends on the deformation potential of the crystal $\hat{\Xi}_{ij}$ ($i$, $j$ representing each of the three Cartesian directions) and the strain tensor components $\hat{U}_{ij}$ \cite{Ridley} (both of which are position dependent in the atomistic TB method), and is given by 
\vspace{-0.15cm}
\begin{equation}  
\hat{H}_{ep}=\sum_{i,j} \hat{\Xi}_{ij} \hat{U}_{ij} \label{vgl1}
\end{equation}
\vspace{-0.3cm}

\noindent
To evaluate $\hat{\Xi}_{ij}$, we use the relation $\hat{\Xi}_{ij}=\frac{\partial \hat{H}_{ep}}{{\partial \hat{U}_{ij}}}$, and compute the total change in the electron-phonon Hamiltonian $\Delta \hat H_{ep}$ due to an infinitesimal uniform lattice strain represented by $\Delta \hat U_{ij}=u_{ij}$ (a small arbitrary constant). Since $\Delta \hat H_{ep}$ can be expressed as a change in the electronic TB Hamiltonian $\hat H_e$ under a crystal deformation caused by $u_{ij}$ \cite{Ridley, Ziman}, $\hat{\Xi}_{ij}$ is given as

\vspace{-0.5cm}
\begin{equation}
  \hat{\Xi}_{ij}=\{\hat{H}_{e}(u_{ij})-\hat{H}_{e}(0)\}/{u_{ij}} \label{vgl5}
\end{equation}
 
\noindent
The strain dependent TB Hamiltonian $\hat{H}_{e}(u_{ij})$ expresses the TB matrix elements as functions of inter-atomic bond lengths and distortion angles depending on the relative positions of pairs of atoms in the lattice \cite{Boykin1}. This method of incorporating local strain in the TB Hamiltonian is well-established and has been shown to reproduce various experimental results \cite{Klimeck_ted}. Although we have considered all 6 components of $\hat{\Xi}_{ij}$, we have found the off-diagonal terms to be small for single donor states located near the conduction band.

Furthermore, $\hat{U}_{ij}$ can also be expressed in terms of phonon creation and annhilation operators, $\hat{a}_q^+$ and $\hat{a}_q$ respectively, as 

\vspace{-0.5cm}
\begin{align}\begin{split}
  \hat{U}_{\rm ij}=\frac{1}{2}\sum_{q}(\frac{\hbar}{2V\rho\omega_{q}})^{\frac{1}{2}}i(e_{\rm qi}q_j+e_{\rm qj}q_i)(\hat{a}_q^+  \exp[i(\boldsymbol{q\cdot{r}})]+\\\hat{a}_q\exp[-i(\boldsymbol{q\cdot{r}})]) \label{vgl2}
\end{split}\end{align}

\noindent
where $V$ is the volume of the crystal, $\rho$ the mass density, and $\boldsymbol{e_q}$ the phonon polarization unit vector. Using eq. 2 and 4, the matrix element of $\hat{H}_{ep}$ can be expressed as, 

\begin{align}\begin{split}
\langle{n',n_q+1|\hat{H}_{ep}|n,n_q}\rangle=\frac{1}{2}\sum_{q}(\frac{\hbar}{2V\rho\omega_{q}})^{\frac{1}{2}}\sqrt{n_q+1}\\\sum_{i,j}i(e_{\rm qi}q_j+e_{\rm qj}q_i)\langle{n'|\exp[i(\boldsymbol{q\cdot{r}})] \hat{\Xi}_{\rm ij}|n}\rangle
 \label{vgl3}
\end{split}\end{align}
\vspace{-0.5cm}

In eq. 5, we have used $\rho=$ 2330 kg/$m^3$, while $\omega_q$ is obtained from the electron Zeeman energy $E_z=\hbar\omega_q$. The phonon number $n_q$ satisfies the Boltzmann distribution. We choose a temperature range below 100mK, which guarantees us to be in the low-temperature regime where ($n_q$+1) $\sim 1$ \cite{Morello_single_shot}. Above 1K, ($n_q+1$) $\sim n_q$, and $1/T_1$ varies as $B^4$ \cite{Hasegawa, Roth}. The polarization vector $\boldsymbol{e_q}$ takes into account the three phonon modes: one longitudinal and two transverse. Since the Zeeman splitting energy is very small ($<$ 1 meV), linear and isotropic bulk Si phonon dispersion is assumed. The phonon wavevector $q$ is then evaluated as $\omega_q=v_sq$, where $v_s$ is the speed of sound in Si, and taken to be 8480 m/s for the longitudinal mode and 5860 m/s for transverse modes. The same constants were also used to interpret the experimental data \cite{Morello_single_shot,Holger_Ncomm}. 
While local vibrational modes have been observed for P atoms in Si \cite{local_vibration}, the energy corresponds to the mode frequency is at least an order of magnitude larger than the energy range in interest. The measured $T_1$ values do not deviate from the $B^5$ behavior indicates that the local vibrational modes do not contribute significantly to the spin-relaxation process.

Fig. 2 shows the spin-lattice relaxation rates, $1/T_1$, of a single P donor and a 4P donor cluster as a function of B-field. The red squares are the measured rates for a single donor from Ref. \cite{Morello_single_shot}, whereas the blue triangles are the measured rates for a STM patterned few donor cluster from Ref. \cite{Holger_Ncomm}. Experimentally, the exact number of donors in the cluster was unknown, but estimated to be between 2 and 5 based on STM images. From transport measurements, it was also expected that during the spin readout step at least three electrons occupied the donor cluster, while in total seven charge transitions on the donor cluster were observed \cite{Holger_Ncomm}. 

The red solid line in Fig. 2 represents the calculations performed in this work for a bulk P donor. The calculated rates show a $B^5$ dependence of $1/T_1$, and also yield similar magnitudes of $T_1$ ($\sim$2.5 s at $B=$ 2 $T$) as the experiment. The B field in this calculation is applied along the [110] direction, consistent with the experiment \cite{Morello_single_shot}. To understand the effect of donor number and electron number on $T_1$, we have simulated donor clusters comprising of 2 to 4 donors with various electron occupation. In Fig. 2, we show the results of the 4P cluster with 1, 3 and 5 bound electrons (the green, cyan, and the blue solid lines, respectively). 
While the $B^5$ dependency holds in all cases, the rates vary considerably in magnitude, and increase with the number of electrons. Our calculations show that higher measured relaxation rates of the cluster come from a 5e occupation in a 4P cluster, which is also consistent with the experimental finding of the electron number being $\geq$ 3 \cite{Holger_Ncomm}.

We have intentionally chosen an odd number of electrons because the relaxation between a net 1/2 and -1/2 spin is assumed, which requires an unpaired number of electrons. This is also consistent with the experimental measurements, where no spin read-out signal was observed for alternate electron occupation in the cluster \cite{Holger_Ncomm}. The calculations also reveal the startling fact that if we have only one electron in a 4P cluster, the relaxation rate is actually smaller than the bulk P, and the $T_1$ times can be increased from few seconds to hundreds of seconds. The physical reason that determines the magnitude of the $T_1$ times is also investigated in this work.

\begin{figure}[htbp]
\includegraphics[trim=50 200 10 200, clip, width=\linewidth]{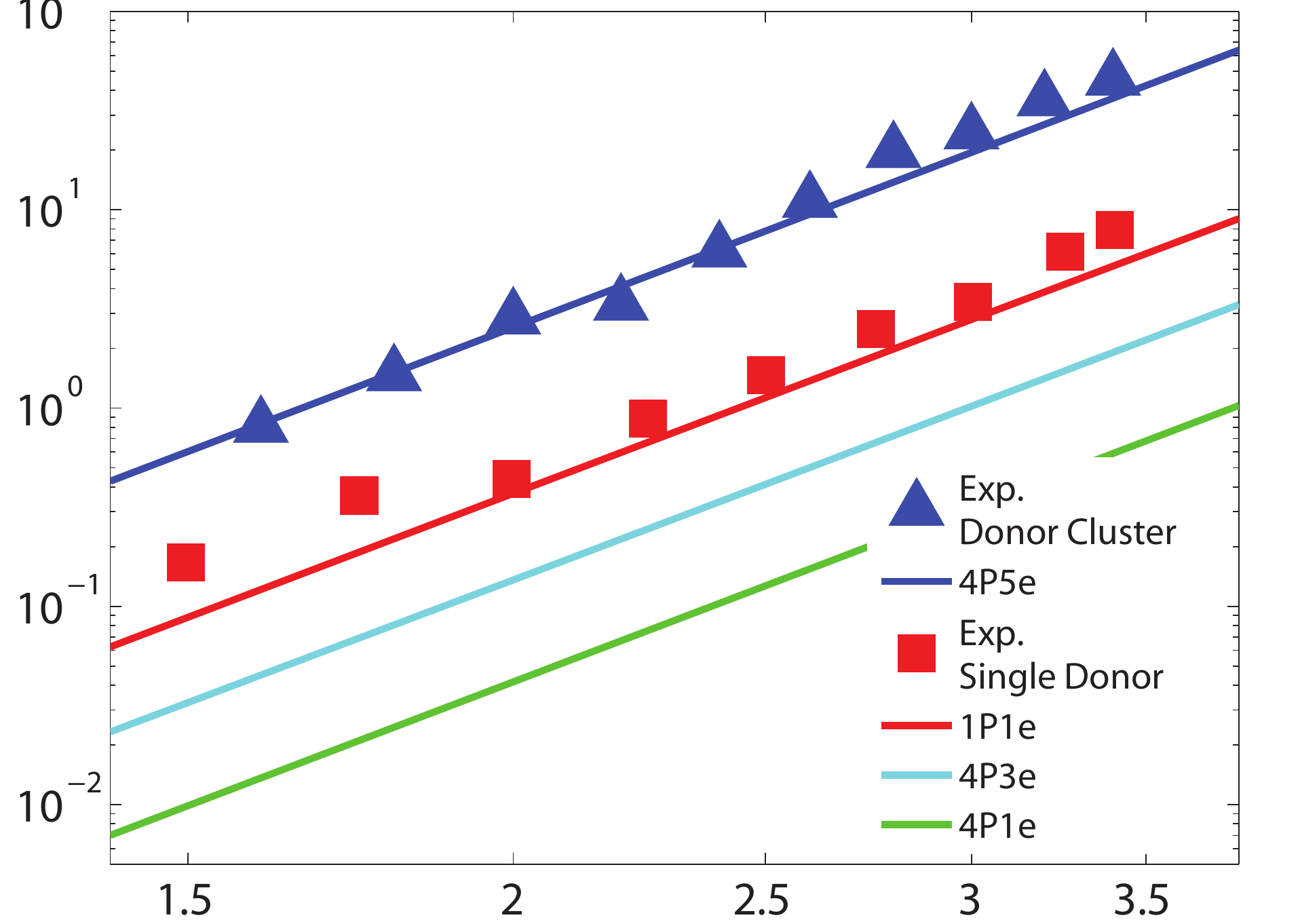}
\caption{ Spin-lattice relaxation rates of a single P donor and a 4P donor cluster as a function of B-field. The red squares and blue triangles show the measured data for a single donor \cite{Morello_single_shot} and a few-donor cluster \cite{Holger_Ncomm}, respectively. The solid lines show the TB calculation results for 1P1e (red), 4P1e (green), 4P3e (cyan) and 4P5e (blue). $1/T_1$ varies as $B^5$ for all cases.}
\vspace{-0.5cm}
\label{fi1}
\end{figure}

\begin{figure}[htbp]
\includegraphics[trim=0 0 0 0, clip, width=\linewidth]{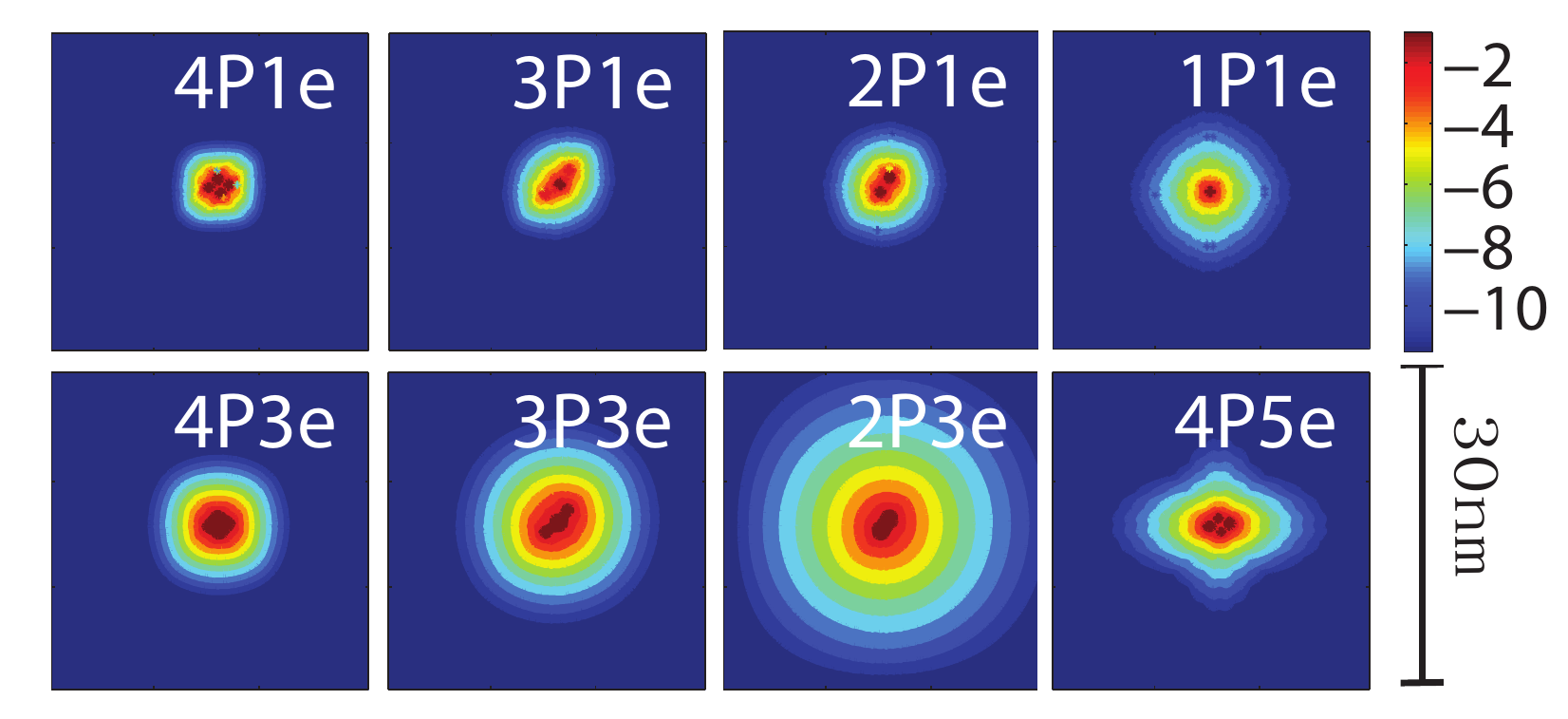}
\caption{The computed probability density (in log scale) of the outermost electron in various donor clusters. The plots show a 2D cut through the center of the clusters.}
\vspace{0cm}
\label{fi2}
\end{figure}

To understand the impact of electron number on the $T_1$ times observed we determined the electron probability densities of the outermost electrons for varying cluster sizes and electron number in Fig. 3. The size and shape of the wavefunctions depends on the number of donors and electrons and their locations within the cluster. For the same number of electrons, more donors result in a more tightly bound wavefunction because of the stronger potential of the larger number of positively charged donor cores. For the same number of donors, as more electrons are added, the wavefunction spreads out more as the donor core becomes more strongly screened. Electron-electron repulsion also causes the wavefunctions of the outermost electron to spread out more. We have extracted the Bohr radii of these wavefunctions by fitting an exponential decay function to the tail of probability densities ($|\Psi(r)|^2$) along the x-axis through the center of the clusters.  

Fig. 4a shows the relaxation rates (at $B=$ 2 T) as a function of the Bohr radii for the same clusters as in Fig. 3. It is observed that donor clusters with larger Bohr radii (i.e., those clusters with more electrons and fewer donors) result in higher relaxation rates. Since the acoustic phonon wavelength corresponding to a Zeeman energy of 0.2 meV (at $B=$ 2 T) is about 100 nm, the phonon wavelength is much larger than the electronic wavelengths in this system. A larger electronic wavefunction therefore interacts with the phonons more \cite{Hanson}. Perhaps more importantly, a larger wavefunction also implies less quantum confinement in the system, and hence a smaller energy gap between excited and ground state (i.e. the valley-orbit gap) \cite{Rahman_VO}, which relates well to the Hasegawa theory for a bulk donor \cite{Hasegawa}, in which $T_1$ increases with this energy gap. Since our calculations show a valley-orbit gap ranging from 30 meV to 5 meV as cluster wavefunction increases in radii, we expect a strong dependence of $T_1$ on radii as well. 
All the calculations presented here also include exchange and correlation effects. We observed that the inclusion of the exchange and correlation effects results in slightly larger wavefunctions, as the electrons experience greater net repulsion. This causes the relaxation rates to increase slightly and move closer to the experimental data in Fig. 2.

\begin{figure}[htbp]
\includegraphics[trim=0 0 0 0, clip, width=\linewidth]{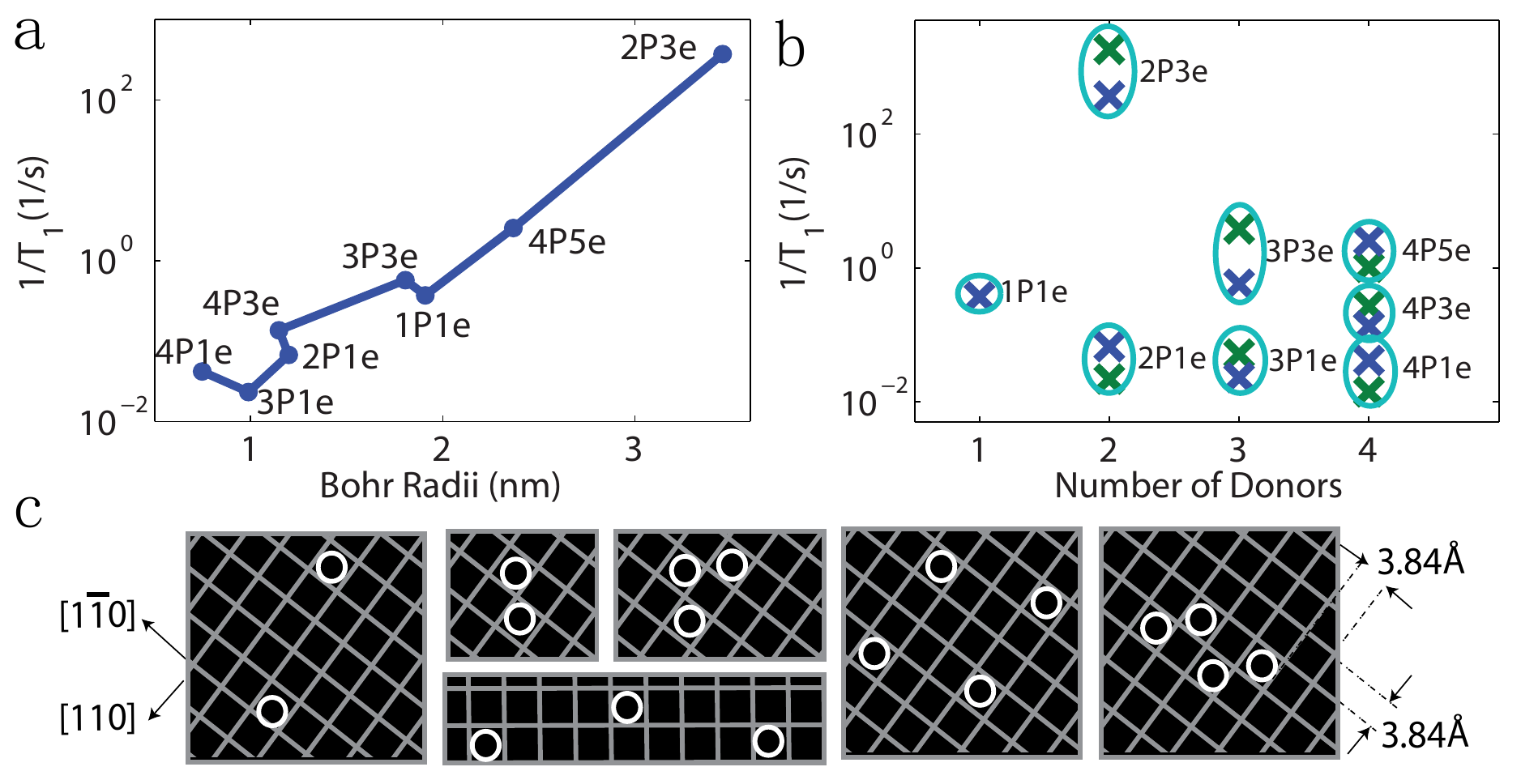}
\caption{(a) The relaxation rates of various donor clusters at $B=$ 2 T as a function of the computed Bohr radii of the wavefunction. (b) The relaxation rates of various donor clusters at $B=$ 2 T with different number of electrons. The different points within an ellipse represent the variations in $T_1$ obtained with different donor locations within the cluster. (c) Exact donor locations for the donor clusters in (b).}  
\vspace{0cm}
\label{fi3}
\end{figure}  

Within a lithographic template for a donor cluster, there can be some uncertainties in the exact locations of the donor \cite{Schofield_PRL}. However, if the cluster only comprises of a few donors, all the positional configurations can be enumerated, and the most compact and the most dispersed donor clusters can be identified. Since it is computationally time intensive to simulate all possible donor configurations within a cluster, we have simulated the two extreme cases for donor clusters of 1 to 4 donors (Fig. 4c).     
In Fig. 4b, we have plotted the relaxation rates as a function of electron number for different donor clusters with these two positional configurations (marked by crosses within an ellipse). As expected, the $T_1$ times have some associated variations with donor locations within a cluster. However, the dependency on the total donor and total electron numbers is stronger. This suggests that $T_1$ measurements can be used to infer information about donor and electron numbers in STM patterned donor devices, as a non-invasive metrology technique. Such information can be useful for engineering pulses to control single- and two-qubit operations in experiments. 

Previous effective mass calculations of Hasegawa \cite{Hasegawa} and Roth \cite{Roth} predicted two different spin relaxation mechanisms due to an effective g-factor shift. Hasegawa's mechanism predicts this effective g-factor shift due to the strain induced redistribution of the donor wavefunction among the six conduction band valleys \cite{Hasegawa}, while Roth's mechanism predicts a single valley g-factor shift due to a strain induced mixing of higher conduction bands \cite{Roth}. Both mechanisms inherently depend on the spin-orbit interaction in silicon, which reduces the bulk donor g-factor slightly below 2. 
Both Hasegawa and Roth's theory show a $B^5$ dependency of $1/T_1$ for a bulk donor, but are rather qualitative in nature. The above approaches are also limited to a single bulk donor for which Kohn-Luttinger wavefunctions can be used \cite{Kohn}. Our TB method captures both mechanisms under the same framework, as a full bandstructure description is used from the atomic orbital basis including spin-orbit interaction. The method is also general and applies to any nanostructures and semiconductors for which accurate TB models can be developed. 

In conclusion, we have presented an atomistic approach to calculate the phonon induced spin lattice relaxation rates in donors in silicon. The $T_1$ times agree very well with recently measured values on single donors and donor clusters, and help to explain the variation of $T_1$ with the numbers of donors and electrons, and the donor locations. The values of $T_1$ were found to have a strong dependency on the size of the electronic wavefunctions. This also provides a way to engineer larger $T_1$ times by using donor clusters with large number of P donors and single electron. 
An atomistic description of the $T_1$ times and their variations in inhomogeneous environment provides crucial information in the design of silicon qubits.

\begin{acknowledgments}
This research was conducted by the Australian Research Council Centre of Excellence for Quantum Computation and Communication Technology (project No. CE110001027), the US National Security Agency and the US Army Research Office under contract No. W911NF-08-1-0527. Computational resources on nanoHUB.org, funded by the NSF grant EEC-0228390, were used. M.Y.S. acknowledges a Laureate Fellowship.
\end{acknowledgments}

Electronic address: hsuehy@purdue.edu, rrahman@purdue.edu

\end{document}